\documentclass[journal=jcp,manuscript=note]{achemso}

\usepackage[version=3]{mhchem} 
\usepackage{graphicx}
\usepackage{amsmath}
\usepackage{amssymb}
\usepackage[singlelinecheck=false]{caption}
\usepackage{array}
\usepackage{xcolor}

\newcolumntype{C}[1]{>{\centering\arraybackslash}p{#1}}

\author{Pavlo Golub}
\author{Pavel Beran}
\author{Andrej Antalik}
\author{Jiri Brabec}
\email{jiri.brabec@jh-inst.cas.cz}

\affiliation[jh-inst]
{J. Heyrovsky Institute of Physical Chemistry, v.v.i., Czech Academy of Sciences, Prague, Czech Republic}

\title[An \textsf{achemso} demo]
  {SC1MC-2022: A database of transition metal complexes for training ML models to predict one-site entropies and mutual information}

\begin{document}

We introduce a new version of the database SC1MC (SC1MC-2022), obtained by extension of the recent SC1MC-2020, which includes artificial mono transition metal complexes. The database involves reference data used as inputs for training of machine learning models, one- and two-site entropies, and mutual information obtained at the DMRG level for canonical and split-localised orbitals. The purpose of this database is to obtain as much as possible information about the electronic correlation structure, which could be exploited by machine learning models to estimate these important information without a significant computational cost for any similar type of systems with some degree of transferability.\\ 

It is hard to over-evaluate the significance of transition metal complexes involved in many biological processes and play crucial role also in industry. For example, they are involved in active sites of enzymes and are responsible for enzyme's catalytic activity, serve as oxygen carriers or are involved in electron transfer reaction as part of proteins, to name a few.
In the same time the complexity of the electronic structure, brought by transition metals, makes the theoretical modeling of such complexes challenging. An overwhelming majority of such complexes can be properly characterized only with very accurate accounting of the dynamical as well as static correlation effects, which is possible only by employing expensive (multireference) methods. \\

Important information about the correlation structure could provide one-site entropies and mutual information (derived from one- and two-orbital entropies). For example, they can be taken as an important criteria for a proper selection of the active space for active space methods (CASSCF, DMRG, etc.),\cite{RH1,ML1}, to optimize ordering of orbitals in active space for DMRG method, to determine bond orders, or to identify metal-insulator transitions.\cite{BT,LS1}

The von Neumann entropy for $N$-site system is defined as

\begin{equation} \label{1p_entrop}
    s^{(N)} = -\sum_{\alpha}{w_{\alpha;1...N} ~ \textrm{ln} (w_{\alpha;1...N})},
\end{equation}

where $w_{\alpha;1,...N}$ are eigenvalues of the reduced $N$-orbital density matrix. The single-site entropy $s^{(1)}$, obtained as a trace over all degrees of freedom but one site, quantify the correlation of the orbital with the rest of the system, i.e. how much quantum information it shares with the remainder of set of orbitals. If $s^{(1)}$ is small, the orbital is weakly correlated with the others and it indicates that the orbital should not be involved in the active space. Of course, this is true only if the active space already involves some important orbitals, which is usually true if the initial orbitals are pre-selected with respect to orbital energies. 

Two-orbital entropy $s^{(2)}$ quantify the correlation for individual pairs of orbitals, i.e. how much a specific pair of orbitals is correlated with the rest of orbitals. The mutual correlation between two orbitals $i$ and $j$ can be obtained from one- and two-orbital entropies $I_{ij}$ (\textit{mutual information}):\cite{Rissler-Noack-2006}

\begin{equation} \label{orbs_entang}
I_{ij} = (s^{(1)}_{i} + s^{(1)}_{j} - s^{(2)}_{ij})(1-\delta_{ij}),
\end{equation}

where the Kronecker function $\delta$ is introduced to make the entanglement 0 in the case $i=j$.\\


\begin{figure}[H]
\includegraphics[width=5cm]{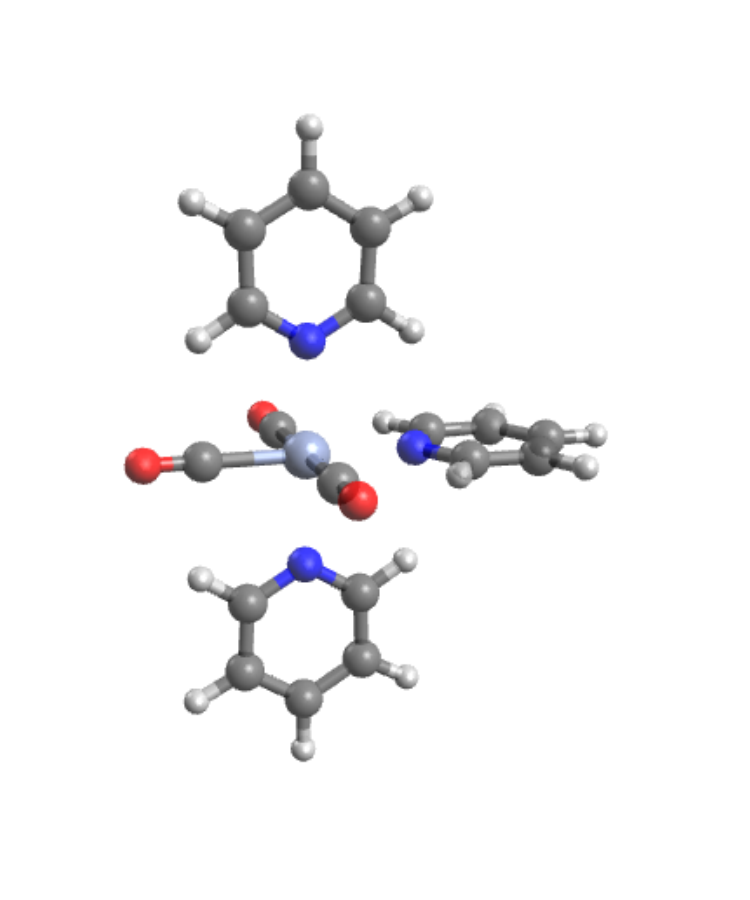}
\caption{An example of the systems from the SC1MC-2022 database}
\label{fig:example}
\centering
\end{figure}

Transition metal complexes have been built by combining seven metals -- Cr, Mn, Fe, Co, Ni, Cu, Zn (the last is not considered as transition metal, however, we decided to include it anyway to generalize the data) -- with a set of ligands (Br\textsuperscript{--}, Cl\textsuperscript{--}, F\textsuperscript{--}, S\textsuperscript{2--}, NH\textsubscript{3}, CO, H\textsubscript{2}O, NC\textsuperscript{--}, N(CH)\textsubscript{5}). The equilibrium metal-ligand distance \textit{x\textsubscript{eq}} has been obtained from geometry optimization at HF level. In the cases when the optimization led to dissociation, \textit{x\textsubscript{eq}} has been set arbitrarily. Complexes with \textit{x\textsubscript{eq}}, 1.5\textit{x\textsubscript{eq}} and 2\textit{x\textsubscript{eq}} metal-ligand distances have been included to the data set. 
We involved complexes with two, four and eight ligands with the limitation that only two different types of ligands can be in one complex. Also we did not mix different metal-ligand distances, i.e. in single complex all ligands can be only at one type of distance from available three types \textit{x\textsubscript{eq}}, 1.5\textit{x\textsubscript{eq}}, 2\textit{x\textsubscript{eq}}. An example of systems is on Fig.\ref{fig:example}.\\

We have generated totally 7,371 complexes, from which 7,259 have converged successfully. For each complex an active space of 36 orbitals has been chosen, where the number of active electrons has been determined from the sum of valence electrons of constituent metal centres and ligands. Assuming that a single orbital is a single sample in machine learning model, this provided in total 261,324 orbital samples. Concerning the two-orbital entropies a pair of orbitals makes a single sample in machine learning models, therefore for this case the dataset provides 4,573,170 samples. \\

Additionally, a part of the data set (with canonical orbitals) has been taken to generate a subset of split-localized samples (occupied and virtual orbitals in the active space are localized separately). For this purpose Foster-Boys localization procedure\cite{FB} has been used. We also added N(CH)\textsubscript{5}) ligand. Active spaces for complexes with more than three piridines have been extended to 46 orbitals by extension of the virtual part. Entropies and mutual information have been recalculated for the new localized spaces. In total 1,035 complexes have been successfully converged, giving additional 971,550 samples for two-orbital entropies.\\

Roughly optimized structures for all complexes have been obtained at Hartree-Fock level with NWChem ab initio software package\cite{Valiev-Bylaska-2010} and Ahlrichs pVDZ basis set\cite{Schafer-Horn-1992}. One- and two-orbital entropies ($s^{(1)}$ and $s^{(2)}$) have been obtained from DMRG\cite{White-1992, White-Martin-1999, Rissler-Noack-2006} calculations with bond dimension 1024 in 16 sweeps. The respective DMRG calculations have been performed with MOLMPS program\cite{MOLMPS}, interfaced with NWChem multiconfiguration SCF (MCSCF) module to read input data including electron integrals directly.\\

The performance of the trained ML models is shown in our recent papers.\cite{ML1,ML2}\\

The information in database is kept in JSON format and includes geometries of the complexes, Hartree-Fock and DMRG total energies, orbital properties for 36 molecular orbitals in the active space , one- and two-electron integrals (one- and two-centre), one- and two-orbital entropies. 

\begin{figure}[H]
\includegraphics[width=14cm]{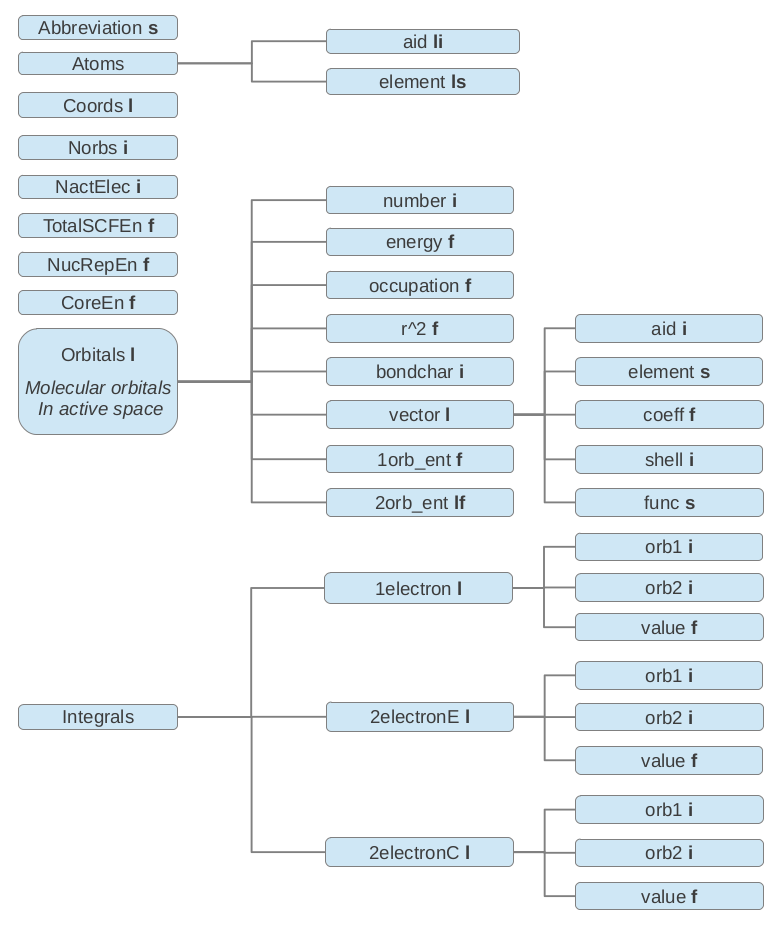}
\caption{The schematic representation of the JSON file interface. The names in boxes correspond to keywords, labels after the names correspond to data type that pairs with the keyword: \textbf{l} -- list, \textbf{i} -- integer, \textbf{f} -- float, \textbf{s} -- string.}
\label{fig:interface}
\centering
\end{figure}

The JSON interface is schematically shown at Figure \ref{fig:interface}. The explanation for keywords:

$\bullet$ \texttt{Abbreviation} -- the abbreviation for molecular complex;

$\bullet$ \texttt{Atoms $\rightarrow$ aid} -- integer identifier for each atom in present molecule;

$\bullet$ \texttt{Atoms $\rightarrow$ element} -- atomic symbol for each atom in present molecule;

$\bullet$ \texttt{Coords} -- Cartesian coordinates for each atom in present molecule, a.u.;

$\bullet$ \texttt{Norbs} -- number of molecular orbitals in active space;

$\bullet$ \texttt{NActElec} -- number of active electrons;

$\bullet$ \texttt{TotalSCFEn} -- converged total energy, a.u.;

$\bullet$ \texttt{NucRepEn} -- nuclear repulsion energy, a.u.;

$\bullet$ \texttt{CoreEn} -- core energy, a.u.;

$\bullet$ \texttt{Orbitals $\rightarrow$ number} -- the molecular orbital number in current active space;

$\bullet$ \texttt{Orbitals $\rightarrow$ energy} -- orbital energy, a.u.;

$\bullet$ \texttt{Orbitals $\rightarrow$ occupation} -- occupation of the orbital;

$\bullet$ \texttt{Orbitals $\rightarrow$ r\^{}2} -- spatial extension of the orbital;

$\bullet$ \texttt{Orbitals $\rightarrow$ bondchar} -- bonding/antibonding character of the orbital;

$\bullet$ \texttt{Orbitals $\rightarrow$ vector} -- set of atomic orbitals from which the molecular orbital is constructed;

$\bullet$ \texttt{Orbitals $\rightarrow$ vector $\rightarrow$ aid} -- integer identifier for atom, which contribute the atomic orbital;

$\bullet$ \texttt{Orbitals $\rightarrow$ vector $\rightarrow$ element} -- atomic symbol for atom, which contribute the atomic orbital;

$\bullet$ \texttt{Orbitals $\rightarrow$ vector $\rightarrow$ coeff} -- expansion coefficient;

$\bullet$ \texttt{Orbitals $\rightarrow$ vector $\rightarrow$ shell} -- atomic shell;

$\bullet$ \texttt{Orbitals $\rightarrow$ vector $\rightarrow$ func} -- function type, e.g 's', 'p\textsubscript{x}', 'p\textsubscript{y}', 'p\textsubscript{z}', 'd\textsubscript{xy}' etc;

$\bullet$ \texttt{Orbitals $\rightarrow$ 1orb\_ent} -- one-orbital entropy;

$\bullet$ \texttt{Orbitals $\rightarrow$ 2orb\_ent} -- two-orbital entropies between present orbital and the rest in active space;

$\bullet$ \texttt{Integrals $\rightarrow$ 1electron} -- list of 1-electron 1,2-center integrals;

$\bullet$ \texttt{Integrals $\rightarrow$ 2electronE} -- list of 2-electron 1,2-center exchange integrals;

$\bullet$ \texttt{Integrals $\rightarrow$ 2electronC} -- list of 2-electron 1,2-center Coulomb integrals.

The distribution of one- and two-orbital entropies by absolute value are presented in tables 1 and 2 respectively. Typically orbitals with one-orbital entropy lesser than 0.05 are regarded as weak. They account for approximately half of all samples in the dataset.

\begin{table}[h!]
\centering
\caption{The distribution of one-orbital entropies by absolute value.}
\label{table: dist1}
\resizebox{\textwidth}{!}{
\begin{tabular}  {C{0.15\textwidth} C{0.25\textwidth} C{0.15\textwidth} C{0.25\textwidth} } 
\hline
\textbf{Value Range} & \textbf{Number of values in range} & \textbf{Value Range} & \textbf{Number of values in range} \\
\hline
$\leqslant$~0.05 & 127~128 & $>$~0.55~$\leqslant$~0.60  &  1~622 \\
\hline
$>$~0.05~$\leqslant$~0.10 & 36~731 & $>$~0.60~$\leqslant$~0.65  &  1~709 \\
\hline
$>$~0.10~$\leqslant$~0.15 & 25~312 & $>$~0.65~$\leqslant$~0.70  &  2~851 \\
\hline
$>$~0.15~$\leqslant$~0.20 & 22~539 & $>$~0.70~$\leqslant$~0.75  &  3~022 \\
\hline
$>$~0.20~$\leqslant$~0.25 & 10~622 & $>$~0.75~$\leqslant$~0.80  &  2~247 \\
\hline
$>$~0.25~$\leqslant$~0.30 & 4~642 & $>$~0.80~$\leqslant$~0.85  &  1~724 \\
\hline
$>$~0.30~$\leqslant$~0.35 & 2~801 & $>$~0.85~$\leqslant$~0.90  &  1~414 \\
\hline
$>$~0.35~$\leqslant$~0.40 & 2~172 & $>$~0.90~$\leqslant$~0.95  &  1~316 \\
\hline
$>$~0.40~$\leqslant$~0.45 & 1~824 & $>$~0.95~$\leqslant$~1.00  &  1~286 \\
\hline
$>$~0.45~$\leqslant$~0.50 & 1~590 & $>$~1.00  &  7~104 \\
\hline
$>$~0.50~$\leqslant$~0.55 & 1~665 & ~  &  ~ \\
\hline
\end{tabular}}
\end{table}

\begin{table}[h!]
\centering
\caption{The distribution of two-orbital entropies by absolute value.}
\label{table: dist2}
\resizebox{\textwidth}{!}{
\begin{tabular}  {C{0.15\textwidth} C{0.15\textwidth} C{0.15\textwidth} C{0.15\textwidth} C{0.15\textwidth} C{0.15\textwidth} } 
\hline
\textbf{Value Range} & \textbf{Number of values in range} & \textbf{Value Range} & \textbf{Number of values in range} & \textbf{Value Range} & \textbf{Number of values in range} \\
\hline
$\leqslant$~0.05 & 911~200 & $>$~0.55~$\leqslant$~0.60  &  50~627 & $>$~1.10~$\leqslant$~1.15  &  42~574\\
\hline
$>$~0.05~$\leqslant$~0.10 & 667~461 & $>$~0.60~$\leqslant$~0.65  &  48~885 & $>$~1.15~$\leqslant$~1.20  &  33~450\\
\hline
$>$~0.10~$\leqslant$~0.15 & 496~981 & $>$~0.65~$\leqslant$~0.70  &  60~999 & $>$~1.20~$\leqslant$~1.25  &  28~650\\
\hline
$>$~0.15~$\leqslant$~0.20 & 504~546 & $>$~0.70~$\leqslant$~0.75  &  82~000 & $>$~1.25~$\leqslant$~1.30  &  23~728\\
\hline
$>$~0.20~$\leqslant$~0.25 & 396~733 & $>$~0.75~$\leqslant$~0.80  &  71~671 & $>$~1.30~$\leqslant$~1.35  &  21~501\\
\hline
$>$~0.25~$\leqslant$~0.30 & 243~389 & $>$~0.80~$\leqslant$~0.85  &  63~667 & $>$~1.35~$\leqslant$~1.40  &  25~066\\
\hline
$>$~0.30~$\leqslant$~0.35 & 173~041 & $>$~0.85~$\leqslant$~0.90  &  55~797 & $>$~1.40~$\leqslant$~1.45  &  14~581\\
\hline
$>$~0.35~$\leqslant$~0.40 & 121~647 & $>$~0.90~$\leqslant$~0.95  &  49~368 & $>$~1.45~$\leqslant$~1.50  &  10~221\\
\hline
$>$~0.40~$\leqslant$~0.45 & 81~569 & $>$~0.95~$\leqslant$~1.00  &  46~659 & $>$~1.50~$\leqslant$~1.55  &  8~032\\
\hline
$>$~0.45~$\leqslant$~0.50 & 59~187 & $>$~1.00~$\leqslant$~1.05  &  45~334 & $>$~1.55~$\leqslant$~1.60  &  6~432\\
\hline
$>$~0.50~$\leqslant$~0.55 & 52~751 & $>$~1.05~$\leqslant$~1.10  &  46~330 & $>$~1.60 &  29~093\\
\hline
\end{tabular}}
\end{table}

The database can be downloaded at https://github.com/brabecj/sc1mc-2022.git

\begin{acknowledgement}

This work has been supported by the Czech Science Foundation (grant no. 19-13126Y) and the Center for Scalable and Predictive methods for Excitation and Correlated phenomena (SPEC), which is funded by the U.S. Department of Energy (DOE), Office of Science, Office of Basic Energy Sciences, the Division of Chemical Sciences, Geosciences, and Biosciences. This work was also supported by the Ministry of Education, Youth and Sports of the Czech Republic through the e-INFRA CZ (ID:90140).

\end{acknowledgement}


\begin{thebibliography}{10}

\bibitem{Jeong-Stoneburner-2020}
Jeong, W.; Stoneburner, S. J.; King, D.; Li, R.; Walker, A.; Lindh, R.; Gagliardi, L.
Automation of Active Space Selection for Multirefernce Methods via Machine Learning
on Chemical Bond Dissociation. 
\textit{J. Chem. Theory Comput.} \textbf{2020}, \textit{16}, 2389 -- 2399.

\bibitem{White-1992}
White, S. R.
Density Matrix Formulation for Quantum Renormalization Groups.
\textit{Phys. Rev. Lett.} \textbf{1992}, \textit{69}, 2863 -- 2866.

\bibitem{White-Martin-1999}
White, S. R.; Martin, R. L.
Ab Initio Quantum Chemistry Using the Density Matrix Renormalization Group.
\textit{J. Chem. Phys.} \textbf{1999}, \textit{110}, 4127 -- 4130.

\bibitem{Rissler-Noack-2006}
Rissler, J.; Noack, R. M.; White, S. R.
Measuring Orbital Interaction Using Quantum Information Theory.
\textit{Chem. Phys.} \textbf{2006}, \textit{323}, 519 -- 531.

\bibitem{FB}
Foster, J. M.; Boys, S. F.
Canonical Configurational Interaction Procedure
\textit{Rev. Mod. Phys.} \textbf{1960}, 32, 300
https://doi.org/10.1103/RevModPhys.32.300

\bibitem{Valiev-Bylaska-2010}
Valiev, M.; Bylaska, E. J.; Govind, N.; Kowalski, K.; Straatsma, T. P.; Dam, H. J.
J. V.; Wang, D.; Nieplocha, J.; Apra, E.; Windus, T. L.; de Jong, W. A.
NWChem: A Comprehensive and Scalable Open-Source Solution for the Large Scale Molecular
Simulation.
\textit{Comput. Phys. Commun.} \textbf{2010}, \textit{181}, 1477 -- 1489.

\bibitem{Schafer-Horn-1992}
Sch\"{a}fer, A.; Horn, H.; Ahlrichs, R.
Fully Optimized Contracted Gaussian Basis Sets for Atoms Li to Kr.
\textit{J. Chem. Phys.} \textbf{1992}, \textit{97}, 2571 -- 2577.

\bibitem{MOLMPS}
Brabec, J.; Brandejs, J.; Kowalski, K.; Xantheas, S.; Legeza, \"{O}.; Veis, L.
Massively Parallel Quantum Chemical Density Matrix Renormalization Group Method.
\textit{Journal of Computational Chemistry}
\textbf{2020}
https://doi.org/10.1002/jcc.26476

\bibitem{ML1}
Pavlo Golub; Andrej Antalik; Libor Veis; Jiri Brabec
Automatic selection of active spaces for strongly correlated systems using machine learning algorithms.
\textit{Journal of Chemical Theory and Computation}
\textbf{2020}
https://doi.org/10.1021/acs.jctc.1c00235

\bibitem{ML2}
Pavlo Golub; Andrej Antalik; Pavel Beran; Jiri Brabec
Mutual information prediction for strongly correlated systems.
\textit{Chemical Physics Letters}
\textbf{2023}
https://doi.org/10.1016/j.cplett.2023.140297.

\bibitem{LS1}
O. Legeza; J. Solyom 
Optimizing the density-matrix renormalization group method using quantum information entropy.
\textit{Phys. Rev. B} \textbf{2003}, 68, 195116.

\bibitem{BT}
Katharina Boguslawski; Pawel Tecmer
Orbital Entanglement in Quantum Chemistry.
\textit{Int. J. Quantum Chem.} \textbf{2015}, 115, 1289--1295 
DOI: 10.1002/qua.24832

\bibitem{RH1}
Christopher J. Stein; Markus Reiher
Automated Selection of Active Orbital Spaces
\textit{J. Chem. Theory Comput.} \textbf{2016}, 12, 4, 1760
https://doi.org/10.1021/acs.jctc.6b00156


\end{thebibliography}
\end{document}